\documentclass[12pt]{article}
\usepackage{epsf}
\begin{document}
\newcommand {\sheptitle}
{\bf \large Quasi-degenerate neutrino masses with normal and inverted hierarchy}
\newcommand {\shepauthor}
{Ng.K.Francis$^{*}$\footnote{\it{E-mail:} ngkf2010@gmail.com}
 and N. Nimai Singh}


\newcommand{\shepaddress}
{Department of Physics, Gauhati University,Guwahati-781014,India\\
$^*$On lien from Department of Physics, Tezpur
  University,Tezpur,India. \\
 \it{E-mail:}ngkf2010@gmail.com}

\newcommand{\shepabstract}
{The effect of CP-phases on quasi-degenerate Majorana neutrino (QDN)
masses are studied with neutrino mass matrix obeying $\mu$-$\tau$
symmetry for normal hierarchy (NH-QD) and inverted hierarchy
(IH-QD). We further investigate on (i) the prediction of solar 
mixing angle below tri-bimaximal value which is consistent with 
observation, (ii) the  prediction on absolute neutrino masses
consistent with $0\nu\beta\beta$ decay mass parameter $(m_{ee})$ and
cosmological bound on the sum of the three absolute neutrino masses 
$\sum_i m_{i}$. The numerical analysis  is carried out through a
parameterization  of neutrino mass matrices using only two unknown
parameters $(\epsilon,\eta)$ within $\mu$-$\tau$ symmetry. The
results  show  the validity of QDN mass models in both normal and
inverted hierarchical patterns.

\vspace{0.3in}

{\bf Keywords:} QDN models, absolute neutrino masses, CP-phases.\\
{\bf PACS Numbers:} 14.60.Pq; 12.15. Ff; 13.40.Em}
\begin{titlepage}
\begin{center}
{\large{\bf\sheptitle}}
\bigskip \\
\shepauthor
\\
\mbox{}
{\it \shepaddress}
\\
\vspace{.5in}
{\bf Abstract} \bigskip \end{center}\setcounter{page}{0}
\shepabstract
\end{titlepage}


\section{Introduction}
Since the present neutrino oscillation data [1] on neutrino mass
parameters are not sufficient to predict the three absolute neutrino
masses in the case of quasi-degenerate neutrino (QDN) mass models
[2-8], such absolute mass scale is usually taken as input parameter
ranging from 0.1 eV to 0.4 eV in most of the theoretical calculations [9]. 
As the latest cosmological tightest upper bound on the sum of the 
three absolute neutrino mass is $\sum_i m_{i}\leq 0.28$ eV [11],
larger value of neutrino mass $m_{3}\geq 0.1$ eV in QDN models, has
 been disfavoured. The upper bound on $m_{ee}\geq 0.2$ eV  in
 $0\nu\beta\beta$ decay [11] also disfavours larger values of 
neutrino mass eigenvalues  with same CP-parity. Some important 
points for further investigations in QDN models for NH-QD and
IH-QD patterns are the searches for QDN models which
can accomodate lower values of absolute neutrino masses $m_{3}\leq 0.09 $ eV, 
solar mixing angle which is lower than tri-bimaximal mixing
(TBM) [12] and effects of CP-phases on neutrino masses. In this paper,
we introduce a general classification for QDN models based on their CP-parity 
patterns and then parameterize the mass matrix within $\mu$- $\tau$ 
symmetry, and finally numerical calculations are carried out.
\section{\bf Parameterization of neutrino mass matrix}
A general $\mu$-$\tau$ symmetric neutrino mass matrix [13,14] with its
four unknown independent matrix elements, requires at least four

independent equations for realistic numerical solution,
\begin{equation}
m_{LL}=
\left(\begin{array}{ccc}
 m_{11}& m_{12} & m_{12}\\
 m_{12} & m_{22} & m_{23}\\
 m_{12} & m_{23} & m_{22}
\end{array}\right).
\end{equation}
The three mass eigenvalues $m_{i}$ and solar mixing angle $\theta_{12}$,
are given by \\
$ m_{1} =
m_{11}-\sqrt{2}\tan\theta_{12}m_{12},\hspace{.1in}\\
 m_{2}=m_{11}+\sqrt{2}\cot\theta_{12}m_{12}, \hspace{.1in}\\ 
 m_{3}=m_{22}-m_{23}$.
\begin{equation}
\tan2\theta_{12}=\frac{2\sqrt{2}m_{12}}{m_{11}-m_{22}-m_{23}}.
\end{equation}
The observed mass-squared differences are calculated as\\ 
\begin{equation}
\bigtriangleup m^{2}_{12} = m^{2}_{2}-m^{2}_{1} >0, \hspace{.1in}
\bigtriangleup m^{2}_{32} =\left| m^{2}_{3}-m^{2}_{2}\right|.\\
\end{equation}\\ 
In the basis where charged lepton mass matrix is diagonal, we have the
leptonic mixing matrix, $U_{PMNS} = U$,  where 
\begin{equation}
U_{PMNS} =
 \left(\begin{array}{ccc}
 \cos\theta_{12} & \sin\theta_{12} & 0\\
\frac{\sin\theta_{12}}{\sqrt{2}} & \frac{\cos\theta_{12}}{\sqrt{2}} & 
-\frac{1}{\sqrt{2}}\\
\frac{\sin\theta_{12}}{\sqrt{2}} & \frac{\cos\theta_{12}}{\sqrt{2}} &
\frac{1}{\sqrt{2}}
\end{array}\right).
\end{equation}     
The mass parameters $m_{ee}$ in $0\nu\beta\beta$ decay and the sum of the
absolute neutrino masses in WMAP cosmological bound $\sum_i m_{i}$, are
given respectively by,
\begin{equation} 
m_{ee} = 
|m_{1}U^{2}_{e1} + m_{2}U^2_{e2}+ m_{3}U^2_{e3}|, \\\\\\  m_{cosmos}=
m_{1}+m_{2}+m_{3}.
\end{equation}
A general classification for three-fold quasi-degenerate neutrino mass
models [13] with respect to Majorana CP-phases in their three mass
eigenvalues, is adopted here. Diagonalization of left-handed Majorana
neutrino mass matrix $m_{LL}$ in eq.(1) is given by 
$m_{LL} = UDU^{T}$, where U is the diagonalising matrix in eq.(4) and
Diag=D$(m_{1},m_{2}e^{i\alpha}, m_{3}e^{i\beta})$ is the diagonal matrix
with two unknown Majorana phases $(\alpha, \beta)$. In the basis where
charged lepton mass matrix is diagonal, the leptonic mixing matrix is
given by $U= U_{PMNS}$ [14]. We then adopt the following classification
according to their CP-parity patterns in the mass eigenvalues
$m_{i}$ namely Type IA: (+-+) for D=Diag$(m_{1},-m_{2},m_{3})$; Type
IB:(+++) for D=Diag$(m_{1},m_{2},m_{3})$ and Type-IC: for (++-) for
D=Diag$(m_{1},m_{2},-m_{3})$ respectively. We now introduce the
following parameterization for $\mu$-$\tau$ symmetric neutrino mass 
matrices $m_{LL}$ which could satisfy the above classifications [13].
\section {Numerical Analysis and Results}
For numerical computation of absolute neutrino masses, we take the
'following observational data:\\ $\bigtriangleup m^{2}_{12}=\left(m^{2}_{2}
- m^{2}_{1} \right )=7.60 \times 10^{-5} eV^{2}$,\\
$\left|\bigtriangleup m^{2}_{32}\right|=\left|m^{2}_{3}-m^{2}_{2}
\right|=2.40\times10^{-3}eV^{2}$;\\ and define the following parameters
$\phi=\frac{\left|\bigtriangleup m^{2}_{23}\right|}{m^{2}_{3}}$ and
$\psi=\frac{\bigtriangleup m^{2}_{21}}{\left|\bigtriangleup m^{2}_{23}\right|}$,
where $m_{3}$ is the input quantity allowed by the latest cosmological
bound. For NH-QD, the other two mass eigenvalues are estimated as,
$m_{2}=m_{3}\sqrt{1-\phi}$; $m_{1}=m_{3}\sqrt{1-\phi(1+\psi)}$ and for
IH-QD as $m_{2}=m_{3}\sqrt{1+\phi}$; $m_{1}=m_{3}\sqrt{1+\phi(1-\psi)}$. 
For suitable input value of $m_{3}$ one can estimate the numerical
values of $m_{1}$ and $m_{2}$ for both NH-QD and IH-QD cases, using
the observational values of $|\bigtriangleup m^{2}_{23}|$ and  
$\bigtriangleup m^{2}_{21}$.Table-1 gives the calculated numerical
values for two models namely NH-QD and IH-QD for 
$\left|\bigtriangleup m^{2}_{23}\right|=7.60\times10^{-5} eV^{2}$  and 
$\bigtriangleup  m^{2}_{21}=2.40\times10^{-3} eV^{2}$.\\

{\bf Parameterizations}:  In the next step we parameterize the mass matrix
eq.(1) into three types: \\{\bf Type IA}  with D=Diag($m_{1},-m_{2},m_{3}$).
The mass matrix of this type [13,15] can be parameterized using two
parameters  $(\epsilon,\eta)$:
\begin{equation}
m_{LL}=
\left(\begin{array}{ccc}
\epsilon-2\eta & -c\epsilon & -c\epsilon\\
-c\epsilon &\frac{1}{2}-d\eta & -\frac{1}{2}-\eta\\ 
-c\epsilon & -\frac{1}{2}-\eta& \frac{1}{2}-d\eta 
\end{array}\right)m_{3}.
\end{equation} 
This predicts the solar mixing angle,
\begin{equation}
\tan^{2}\theta_{12}=-\frac{2c\sqrt{2}}{1+(d-1)\frac{\eta}{\epsilon}}.
\end{equation}
when we choose the constant parameters c=d=1.0, we get the
tri-bimaximal mixings (TBM) $tan2\theta_{12}=-2\sqrt{2}$ which leads
to $\tan^{2}\theta_{12}= 0.50$ and the values of $\epsilon$ and $\eta$
are calculated for both NH-QD and IH-QD cases, by using the values 
of Table-1 in these two eigenvlue expressions:
$m_{1}=(2\epsilon-2\eta)m_{3}$ and $m_{2}=(-\epsilon-2\eta)m_{3}$
which are extracted after diagonalization of eq.(6). 
The results are given in Table-2 for $tan^{2}\theta_{12}=0.50$. 
The solar angle can be further lowered by taking the values $c<1$ 
and $d<1$ while using the earlier values of $\epsilon$ and $\eta$ 
extracted for TBM case. For $\tan^{2}\theta_{12}=0.45$ case the
results  are shown in Table-3. \\{\bf Type-IB} with 
D = Diag $(m_{1}, m_{2}, m_{3})$: This type [13,15] of
quasi-degenerate mass pattern is given by the mass matrix, 
\begin{equation}
m_{LL} = \left(\begin{array}{ccc}
1-\epsilon-2\eta & c\epsilon & c\epsilon\\ 
c\epsilon &  1-d\eta & -\eta\\ 
c\epsilon & -\eta & 1-d\eta
\end{array} \right)m_{3}. 
\end{equation}  
This predicts the solar mixing angle, 
\begin{equation}
\tan2\theta_{12} =
\frac{2c\sqrt{2}}{1+(1-d)\frac{\eta}{\epsilon}}.                
\end{equation}
which gives the TBM solar mixing angle with the input values c = 1 and
d = 1. When $\epsilon=0$, $\eta=0$, this leads to
$m^{diag}_{LL}=diag(1,1,1)m_{3}$.
Like in Type-IA, here {\bf $\epsilon$} and {\bf $\eta$}
values are computed for NH-QD and IH-QD, by using Table-1 in 
$m_{1}=(1-2\epsilon-2\eta)m_{3}$ and $m_{2} = (1+\epsilon-2\eta)m_{3}$
which are extracted from diagonalization of eq.(8).\\
{\bf Type-IC} with D = Diag($m_{1},m_{2},-m_{3}$): It is not
necessary to treat this model [13] separately as it is similar to
Type-IB except with the interchange of two matrix elements ($m_{22}$)
and $(m_{23})$ in the mass matrix in eq.(10), and this effectively
imparts an additional odd CP-parity on the third mass eigenvalue
$m_{3}$ in Type-IC. Such change does not alter the predictions of
Type-IB. Tables 2 and 3 present our numerical results for both 
$tan^{2}\theta_{12}=0.5$ and 0.45 cases, in all types of QD models
(Types-IA, IB, IC). These results are consistent with observational 
cosmological bound.
\begin{table}
\begin{tabular}{|c|c|c|c|c|c|}  \hline
\emph{input} & \emph{calculated} & 
\multicolumn{2}{c|}{\emph{NH-QD}} &  
\multicolumn{2}{c|}{\emph{IH-QD}}  \\ \cline{3-6}
\bf$m_{3}$ & \bf $\phi$ & \bf $m_{1}$ & \bf $m_{2}$ & \bf $m_{1}$ &\bf $m_{2}$ \\  \hline\hline
0.40 & 0.015 & 0.39689 & 0.39699 & 0.40289 & 0.40299\\
0.10 & 0.24 & 0.08674 & 0.08718 & 0.11104 & 0.11135\\
0.08 & 0.375 & 0.06264 & 0.06325 & 0.09340 & 0.09380\\\hline
\end{tabular}
\caption{The absolute neutrino masses  in eV are estimated from
  oscillation data (using calculated $\psi=0.031667$).}
\end{table}
\begin{table}
\begin{tabular}{|l|l|l||l|l|}  \hline
\emph{Different} & 
\multicolumn{2}{c||}{\emph{NH-QD}} &  
\multicolumn{2}{c|}{\emph{IH-QD}}  \\ \cline{2-5}
\bf{parameters} & \bf Type-IA & \bf Type-IB & \bf Type-IA & \bf
Type-IB \\ \hline\hline

c & 1.0 & 1.0 & 1.0 & 1.0\\ 
d & 1.0 & 1.0 & 1.0 & 1.0\\
$m_{3}$ & 0.10 & 0.10 & 0.08 & 0.08\\
$\epsilon$ & 0.57972 & 0.0015 & 0.78004 & 0.00169\\
$\eta$ & 0.14602 & 0.0649 & 0.19628 & -0.08546\\ \hline

$m_{1}$ (eV) & 0.08674 & 0.08675 & 0.09340 & 0.09340\\
$m_{2}$ (eV) & -0.08717 & 0.08717 & -0.09380 & 0.09380\\
$m_{3}(eV)$ & 0.10 &0.10 & 0.08 & 0.08 \\ 
$\sum|m_{i}|eV$ & 0.27 & 0.274 & 0.267 & 0.274\\ \hline

$\bigtriangleup m^{2}_{21}eV^{2}$ & $7.6\times10^{-5}$ &
$7.6\times10^{-5}$ & $7.6\times10^{-5}$ & $7.6\times10^{-5}$\\
$\left|\bigtriangleup m^{2}_{23} \right| eV^{2}$ & $2.2\times10^{-3}$ &
$2.4\times10^{-3}$ & $2.4\times10^{-3}$ & $2.4\times10^{-3}$\\  
$\tan^{2}\theta_{12}$ & 0.50 & 0.50 & 0.50 & 0.50\\
$|m_{ee}|$ eV  & 0.08688 & 0.0869 & 0.09354 & 0.09354\\ \hline
\end{tabular}
\bf\caption{Predictions for  \bf$\tan\theta_{12}=0.50$}
\end{table}
\begin{table}
\begin{tabular}{|l|l|l||l|l|}  \hline

\emph{Different} & 
\multicolumn{2}{c||} {\emph{NH-QD}} &  

\multicolumn{2}{c|} {\emph{IH-QD}}  \\ \cline{2-5}
\bf {parameters} & \bf Type-IA & \bf Type-IB & \bf Type-IA & \bf Type-IB \\ \hline\hline 

c & 0.868 & 0.945 & 0.868 & 0.96\\
d & 1.025 & 0.998 & 1.0 & 1.002\\
$m_{3}$ & 0.10 & 0.10 & 0.08 & 0.08\\
$\epsilon$ & 0.6616 & 0.00145 & 0.88762 & 0.00169\\
$\eta$ & 0.1655 & 0.06483 & 0.22317 & -0.08546\\ \hline

$m_{1}$ (eV) & 0.0876 & 0.08676 & 0.09392 & 0.09341 \\
$m_{2}$ (eV) & -0.0880 & 0.08717 & -0.09432 & 0.09381\\
$m_{3}(eV)$ & 0.0996 &0.10002 & 0.08 & 0.080014 \\ 
$\sum|m_{i}|eV$ & 0.274 & 0.274 & 0.268 & 0.267\\ \hline

$\bigtriangleup m^{2}_{21}eV^{2}$ & $7.7\times10^{-5}$ &
$7.3\times10^{-5}$ & $7.6\times10^{-5}$ & $7.4\times10^{-5}$\\
$\left|\bigtriangleup m^{2}_{23} \right| eV^{2}$ & $2.2\times10^{-3}$ &
$2.4\times10^{-3}$ & $2.4\times10^{-3}$ & $2.4\times10^{-3}$\\  
$\tan^{2}\theta_{12}$ & 0.45 & 0.45 & 0.45 & 0.45\\
$|m_{ee}|$ eV  & 0.0877 & 0.08688 & 0.09403 & 0.09354\\ \hline
\end{tabular}
\bf\caption{Predictions for \bf$\tan\theta_{12}=0.45$}
\end{table}
\section{Conclusion}
To conclude, we have studied the effects of Majorana phases on the
prediction of absolute neutrino masses in three types of QDN models
having both normal and inverted hierarchical patterns within $\mu$-$\tau$
symmetry. These predictions are consistent with data on the mass
squared difference derived from various oscillation experiments, and
from the upper bound on absolute neutrino masses in $0\nu\beta\beta$
decay as well as upper bound of $\sum_i m_{i}\leq 0.28$ eV  from
cosmology. The QD models are still far from discrimination and the
prediction on solar mixing angle is found to be lower than TBM
viz,$\tan^{2}\theta_{12}=0.45$ which coincides with the best-fit in the
neutrino oscillation data. The result shows the validity of NH-QD and
IH-QD models. The results presented in this article are new  and have
important implications in the discrimination of neutino mass models.

\vspace{.1in}
{\large \bf Acknowledgement}\\
One of us (Ng.K.F) wishes to thank the University Grants Commission,
Govenment of India for sanctioning the project entitled {\bf ``Neutrino
masses and mixing angles in neutrino oscillations''} vide Grant. No
32-64/2006 (SR). This work was carried out through this project.

\end{document}